\documentstyle[preprint,aps]{revtex}
\begin{document}
\draft
\title{Cluster variation -- Pad\'e approximants method for the simple
cubic Ising model}
\author{Alessandro Pelizzola}
\address{Istituto Nazionale per la Fisica della Materia, Unit\`a
Torino Politecnico,  and 
Dipartimento di Fisica del Politecnico di Torino,
c. Duca degli Abruzzi 24, 10129 Torino, Italy}
\date{\today}
\maketitle
\tightenlines
\begin{abstract}
The cluster variation -- Pad\'e approximant method is a recently
proposed tool, based on the extrapolation of low/high temperature
results obtained with the cluster variation method, for the
determination of critical parameters in Ising-like models. Here the
method is applied to the three-dimensional simple cubic
Ising model, and new results, obtained with an 18-site basic cluster,
are reported. Other techniques for extracting non-classical critical
exponents are also applied and their results compared with those by the
cluster variation -- Pad\'e approximant method.  
\end{abstract}


The cluster variation method (CVM) \cite{kik1,an,morita} is a
hierarchy of approximations which generalizes the well-known mean
field approximation and has been widely applied in the last decades,
mainly to study the equilibrium properties of classical, discrete
lattice models with short range interactions. The CVM results are more
and more accurate as the size of the clusters considered increases (at
least in a specific way, see below),
but the critical exponents take always the mean field (classical)
values. The issue of extracting non-classical critical exponents from
mean field approximations has been the subject of a certain amount of
research work in recent years. As far as the CVM is concerned, a few
schemes have been proposed in recent years to give estimates of
critical exponents from the CVM results. One of these schemes, the
cluster variation -- Pad\`e approximant method (CVPAM)
\cite{ap-prerc,ap-pre}, was specifically devised for the CVM and
exploits its great accuracy at high and low temperatures by means of
an extrapolation of the thermodynamic quantities based on Pad\`e
approximants.

In the present paper I report on an
investigation on the CVM approximation for the Ising model on the
simple cubic lattice with the largest basic cluster (18 sites) ever
considered. The results of this approximation are used to give
non-classical estimates of the three-dimensional Ising critical
exponents, using mainly the cluster variation -- Pad\`e approximant
method (CVPAM). Other schemes, like the
coherent anomaly method (CAM) \cite{CAMbook} and an approach by Tom\'e
and de Oliveira \cite{tome} are also considered.

I shall study the Ising model on the simple cubic lattice, with
nearest-neighbor (NN) interactions only, described by the reduced
Hamiltonian
\begin{equation}
\frac{\cal H}{k_B T} = - K \sum_{\langle i j \rangle} s_i s_j,
\end{equation}
where $K$ is the (reduced) interaction energy, $s_i = \pm 1$ is the
usual Ising variable at site $i$, the summation runs on NN pairs, and
$k_B$ and $T$ are, as customary, Boltzmann's constant and absolute
temperature, respectively. 

The CVM is a variational method based on the minimization of a free
energy density which is obtained \cite{an,morita} by truncating the
cumulant expansion of the exact variational principle of equilibrium
statistical mechanics. The approximate free energy density depends on
the density matrix (matrices) of the largest cluster(s) entering the
expansion, which completely determine the approximation. The simple
cubic Ising model has been investigated, using the CVM, by Kikuchi
\cite{kik1} in the NN pair, square and cube approximations and by the
present author \cite{ap-physa} in the star-cube approximation. Here I
shall use the 18-point approximation which is obtained by choosing as
basic cluster the $3 \times 3 \times 2$ cluster obtained by joining
four cubes as in Fig.\ \ref{BasicCluster}, which is the largest
cluster ever considered for this lattice. The choice of this cluster
is motivated by Schlijper's observation \cite{schlijper} that, due to
the existence of a transfer matrix for the model, accuracy can be
increased by enlarging the basic clusters in $d-1$ dimensions only,
where $d$ is the lattice dimensionality. One could imagine a series
(the generalization of the so-called C-series by Kikuchi and Brush
\cite{brush}) of basic clusters made of $L \times L \times 2$ sites,
where for $L = 2$ one has the cube approximation, for $L = 3$ the
present one, and for $L > 3$ approximations which cannot be dealt with
using current computers. 

Following An \cite{an}, the (reduced) free
energy density to be minimized can be written in the form
\begin{eqnarray}
f(\rho_{18}) = -3 K {\rm Tr} (s_1 s_2 \rho_4) &&+ {\rm Tr} (\rho_{18} \ln
\rho_{18}) - 2 {\rm Tr} (\rho_{12} \ln \rho_{12}) - {\rm Tr} (\rho_{9} \ln
\rho_{9}) \nonumber \\
&& + {\rm Tr} (\rho_{8} \ln \rho_{8}) + 2 {\rm Tr} (\rho_{6} \ln
\rho_{6}) - {\rm Tr} (\rho_{4} \ln \rho_{4}),
\end{eqnarray}
where $\rho_n$ denotes the density matrix of the $n$-point cluster
(see Fig.\ \ref{BasicCluster}), while ${\rm Tr} (s_1 s_2 \rho_4)$ is
the NN correlation ($s_1$ and $s_2$ being any NN sites on the square
plaquette). 

The free energy density can be regarded as dependent on
$\rho_{18}$ only, because the subcluster density matrices can be
defined as suitable partial traces of $\rho_{18}$. Since the
hamiltonian is classical, the density matrices 
are diagonal, and hence our free energy density depends, in principle,
on the $2^{18}$ diagonal
elements of $\rho_{18}$ corresponding to the spin
configurations of the 18-point cluster. These elements are, however,
not all independent. First of all, the density matrices must be
normalized according to ${\rm Tr} \rho_{18} = 1$. In addition, many
elements are degenerate because of the lattice symmetries. The
nondegenerate $\rho_{18}$ elements turn out to be 17676, a rather
small number if compared with $2^{18} = 262144$. Taking into account
lattice symmetries the size of the problem is reduced by a factor
slightly smaller than 16, which is the number of elements of the
symmetry group of our 18-point basic cluster. Finally, it must be
observed that $\rho_{12}$ can be defined in different ways as a
partial trace of $\rho_{18}$. In order to ensure that these different
traces yield the same density matrix one has to impose 1134
constraints on the elements of $\rho_{18}$. Once these constraints are
satisfied no ambiguity is left in the definition of the other
subcluster density matrices. Therefore, one is left with the problem
of finding the minimum of a function of 17676 variables, with 1134
(leaving apart the trivial normalization constraint) linear
constraints among them. The problem can be easily treated in the
framework of the natural iteration method (NIM) \cite{kik2,kik3},
developed by Kikuchi for the solution of the CVM variational
problem. The solution for a single value of $K$ (not too close to the
critical point) can be found on a modern personal computer in a time
of the order of ten minutes.

The accuracy of the present approximation can be assessed in several
ways. For instance, I have compared in Tab.\ \ref{Kc} the present
$K_c$ value with those from other CVM approximations and with the best
estimates \cite{blote,baillie,guttmann,gupta,talapov}. Another
interesting check is the comparison, in Tab.\ \ref{mvsT}, of our
magnetization values $m = \langle s_i \rangle$ with those given by the
formula by Talapov and Bl\"ote \cite{talapov}, determined on the basis
of high precision simulations and finite size scaling. It is
interesting to observe that the best agreement between the two methods
occurs in the middle of the temperature range considered (which lies
within the temperature range $t = 1 - K_c/K \in (0.0005;0.26)$, where
Talapov and Bl\"ote regard their result as very accurate). For $T =
1/K = 3.7$ the two magnetizations differ only by $10^{-7}$. For larger
temperatures our results are certainly less accurate than those by
Talapov and Bl\"ote, while the inverse must be true for smaller
temperatures (the result by Talapov and Bl\"ote is significantly
smaller than 1 at very low temperatures and has a maximum around $T =
1.85$).

In order to obtain non-classical estimates of the critical exponents,
I shall now apply the CVPAM according to the rules outlined in
\cite{ap-prerc,ap-pre}. In the CVPAM one computes a thermodynamical
function $F$ for a set of temperature values in a range where the CVM
approximation can be regarded as very accurate. These values are then
used as a basis for extrapolation by means of Pad\`e approximants and
their generalizations. 

As a first step I shall consider the low
temperature magnetization as a function of the variable $x =
e^{-K}$. In order to determine a temperature range $x < x_{\rm max}$
in which the 18-point CVM is very accurate, I compare it with a lower
order CVM approximation. Although the star-cube approximation is
slightly more accurate, I choose the cube approximation for this
purpose, since, as I mentioned above, the cube and 18-point
approximations can be 
thought of as belonging to the same series. Requiring that the
magnetization difference is less than the empirically determined
\cite{ap-pre} threshold $\epsilon = 10^{-5}$ I obtain $x_{\rm max} =
0.75$. The function $m(x)$ has now to be extrapolated to estimate its
singular properties, taking into account also confluent singularities,
that is corrections to scaling. To this end I use Adler's
generalizations \cite{adler1,adler2,adler3}, usually denoted by M1 and
M2, of the ordinary Pad\`e approximant method (I recall that an
$[L,M]$ Pad\`e approximant is simply the ratio of two polynomials of
degree $L$ and $M$ \cite{domb3,domb13}). Given a function $F(x)$
with a singularity which can be assumed of the form $(x_c -
x)^{-\lambda} [1 + a(x_c - x)^{\Delta_1}]$ method M1 considers Pad\`e
approximants to the logarithmic derivative of the function 
\begin{equation}
B(x) = \lambda F(x) - (x_c - x) \frac{dF}{dx}
\end{equation}
for assigned $x_c$ and $\lambda$. The dominant singularity in $(d/dx)
\ln B(x)$ is a pole at $x_c$ with residue $\lambda - \Delta_1$ if
$\Delta_1 < 1$ and $\lambda - 1$ otherwise. Method M2 considers
instead, for assigned $x_c$ and $\Delta_1$, Pad\`e approximants to 
\begin{equation}
G(y) = - \Delta_1 (y-1) \frac{d\ln F}{dy}, \qquad 
y = 1 - \left(1 - \frac{x}{x_c}\right)^{\Delta_1},
\end{equation}
which should converge to $\lambda$ for $y = 1$. In the CVPAM, the
function $d\ln B/dx$ or $G(y)$ is evaluated at $L + M + 1$ equally
spaced points $x_n = x_{\rm max} - n \delta x$, $x = 0, 1, \ldots L+M$
and then an $[L,M]$ Pad\`e approximant is determined by
interpolation. The value of the spacing $\delta x$ must be adjusted
empirically so that the sets of linear equations which must be solved for the
interpolation are not badly conditioned. For the magnetization, the
best conditioned sets of equations are obtained for $\delta x$ = 
0.015. 

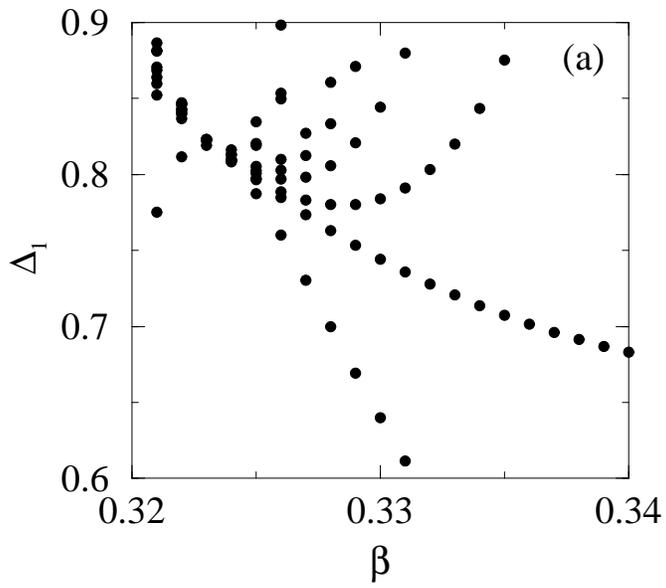
\begin{figure}
\caption{$\Delta_1$ vs $\beta$ plot obtained by method M1, for $T_c =
4.512$ (a), 4.513 (b), 4.514(c) and 4.515 (d), using $[L,M]$
approximants with $3 \le L \le 5$, $L-1 \le M \le L+1$.}
\label{M1}
\end{figure}

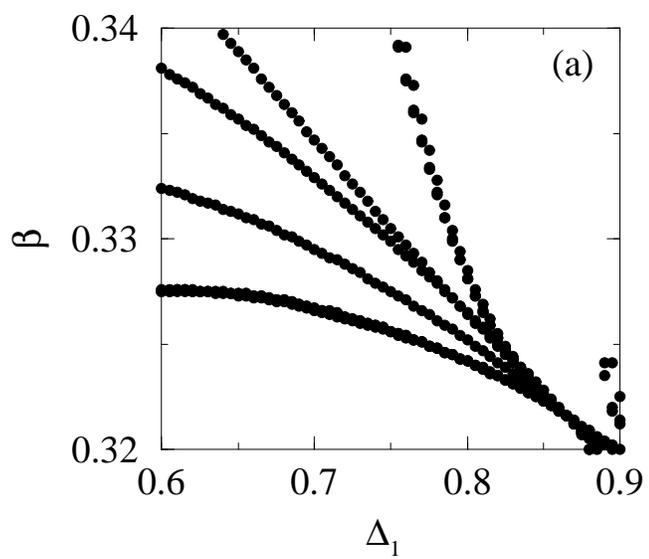
\begin{figure}
\caption{$\beta$ vs $\Delta_1$ plot obtained by method M2, for $T_c =
4.512$ (a), 4.513 (b), 4.514(c) and 4.515 (d), using $[L,M]$
approximants with $3 \le L \le 5$, $L-1 \le M \le L+1$.}
\label{M2}
\end{figure}

Applying method M1 to our magnetization estimates we have obtained the
correction to scaling exponent $\Delta_1$ as a function of the
critical exponent $\beta$ for several values of the trial critical
temperature $T_c = 1/K_c$. It is known \cite{adler3} that the plots
given by M1 have a different curvature above and below the critical
temperature. In the ordered phase the $\Delta_1$ versus $\beta$ plot
is bent upward, while in the disordered phase it is bent
downward. Using this criterion we can locate the critical temperature
in the range $4.512 \le T_c \le 4.515$, which corresponds to $0.22148
\le K_c \le 0.22163$, in agreement with the most recent estimates
\cite{blote,baillie,guttmann,gupta,talapov}. From the corresponding
plots, reported in Fig.\ \ref{M1}, one reads the estimates $0.323 <
\beta < 0.332$ for the magnetization critical exponent and $0.75 <
\Delta_1 < 0.82$ for the correction to scaling exponent. As in the
case of the face-centered cubic lattice \cite{ap-pre}, the critical
exponent result is consistent with recent estimates
\cite{blote,talapov}, while the correction to scaling one is
substantially higher. Method M2 does not provide a clear-cut way to
estimate $T_c$, but using the result by M1 as an input we get (see
Fig.\ \ref{M2}) $0.322 < \beta < 0.331$ and $0.79 < \Delta_1 < 0.87$.

I tried to analyze the high temperature susceptibility data (which
typically give much better results than the low temperature one) in
the same way. I considered the susceptibility as a function of the
variable $w = {\rm tanh} K$ and determined a region $w < w_{\rm max}$
of the disordered phase in which the 18-point CVM approximation can be
regarded as very accurate by comparing the NN correlation (which is
bounded, and hence is better than the susceptibility for such a
test) with that given by the cube approximation. The two estimates
differs by less than $\epsilon = 10^{-5}$ when $w < w_{\rm max} =
0.13$. Unfortunately, both methods M1 and M2 failed to converge and I
had to resort to ordinary Pad\`e approximants for the logarithmic
derivative of the susceptibility. The points for the interpolation
were chosen as described previously, with a spacing $\delta w =
0.003$. Results for the critical exponent $\gamma$ from $[L,L]$
approximants biased with the above estimates for $T_c$ are reported in
Tab.\ \ref{gamma}. Including also results from $[L,L \pm 1]$
approximants I can conclude $1.237 < \gamma < 1.248$, which again is
consistent with the most recent estimates. 

Finally, I have tried to analyze the magnetization data by means of
other techniques, namely the CAM \cite{CAMbook} and a similar approach
by Tom\'e and de Oliveira \cite{tome}. 

The CAM scaling hypothesis for the magnetization \cite{CAMbook} is
that its critical amplitude, defined by $m(T) \simeq B^*(T^*-T)^{1/2}$
must diverge as $B^* \sim (T^* -
T_c)^{\beta - 1/2}$, where $T_c$ is the true critical temperature and
$T^*$ and $B^*$ are the estimates for the critical temperature and the
critical amplitude in a given approximation. Using the above mentioned
best estimate $K_c = 0.22165$ for the critical temperature, a fit on
the results from the pair, cube, star-cube and 18-point CVM
approximation (the plaquette approximation was discarded since it was
clearly out of the curve) I got $\beta = 0.351$. Poorer results were
obtained discarding, in addition to the plaquette approximation, the
pair approximation or the star-cube approximation ($\beta = 0.415$ and
0.212 respectively).

The approach by Tom\'e and de Oliveira \cite{tome} is based on the
scaling assumption $m(T_c) \sim (T^* - T_c)^\beta$. The best results
with this approach have been found by fitting the results from the
pair, cube and 18-point approximations ($\beta = 0.312$) and those
from the cube and 18-point approximations only ($\beta = 0.344$). 

In conclusion, I have developed an 18-point (the largest maximal cluster
ever considered) CVM approximation for the simple cubic lattice and
applied it to the Ising model. The results from this approximation
have been used to extract non-classical estimates for the critical
exponents. Among the three methods considered for this purpose, namely
the CVPAM, the CAM and the method by Tom\`e and de Oliveira, the CVPAM
is the only one which gives critical exponents which are (except for the
correction to scaling exponent) consistent with the most recent
estimates. The CVPAM is therefore to be preferred when extrapolating
CVM results to the critical region. The effort needed was essentially
the same in all cases, 
since most of the labor and computer time go into the development and
solving of the CVM approximation. In particular, the computer
resources needed are remarkably small if compared with the
requirements of extensive Monte Carlo simulations or series
expansions, and the results are only slightly poorer, a feature which
makes the CVPAM an interesting technique whenever powerful computers
are not available and/or very high accuracy is not needed.

\mediumtext

\begin{table}
\caption{Comparison of various CVM estimates for the critical point
with the best estimates}
\label{Kc}
\begin{tabular}{ccccccc}
Method & Pair \cite{kik1} & Square \cite{kik1} & Cube \cite{kik1} &
Star-cube \cite{ap-physa} & 18-point (present) & Best estimates 
\cite{blote,baillie,guttmann,gupta,talapov} \\
$K_c$ & 0.20273 & 0.21693 & 0.21829 & 0.2187 & 0.2199 & 0.22165 \\
\end{tabular}
\end{table}

\begin{table}
\caption{Comparison of our estimate of the magnetization with that
by Talapov and Bl\"ote}
\label{mvsT}
\begin{tabular}{ccc}
$T = 1/K$ & $m$ (present work) & $m$ (Talapov and Bl\"ote
\cite{talapov}) \\
\hline
3.4 & 0.8972562 & 0.8972440 \\
3.5 & 0.8806417 & 0.8806366 \\
3.6 & 0.8616750 & 0.8616735 \\
3.7 & 0.8399256 & 0.8399255 \\
3.8 & 0.8148173 & 0.8148161 \\
3.9 & 0.7855490 & 0.7855416 \\
4.0 & 0.7509519 & 0.7509251 \\
4.1 & 0.7092094 & 0.7091249 \\
4.2 & 0.6572414 & 0.6569722 \\
4.3 & 0.5891051 & 0.5881361 \\
4.4 & 0.4905811 & 0.4859045 \\
4.5 & 0.3067063 & 0.2378014 \\
\end{tabular}
\end{table}

\begin{table}
\caption{Critical exponent $\gamma$ from biased $[L,L]$ approximants}
\label{gamma}
\begin{tabular}{cccccc}
& $T_c$ & 4.512 & 4.513 & 4.514 & 4.515 \\
$L$ & & & & & \\
\hline
4 & & 1.24748  & 1.24520  & 1.24296  & 1.24076  \\
5 & & 1.24748  & 1.24519  & 1.24289  & 1.24058  \\
6 & & 1.24748  & 1.24502  & 1.24217  & 1.23895  \\
7 & & 1.24749  & 1.24505  & 1.24228  & 1.23922  \\
8 & & 1.24748  & 1.24482  & 1.24155  & 1.23769  \\
9 & & 1.24749  & 1.24483  & 1.24156  & 1.23769  \\
10 & & 1.24746  & 1.24487  & 1.24207  & 1.23905  \\
\end{tabular}
\end{table}

\begin{figure}
\caption{The 18-point basic cluster and its subclusters}
\label{BasicCluster}
\end{figure}

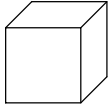
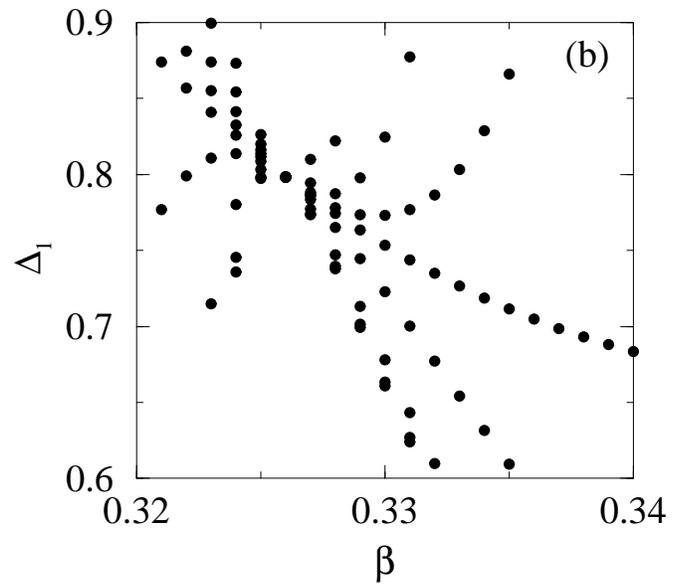
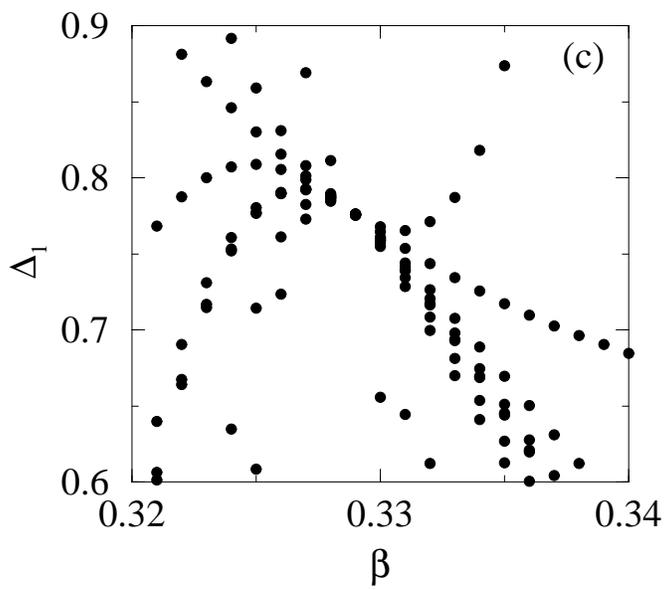
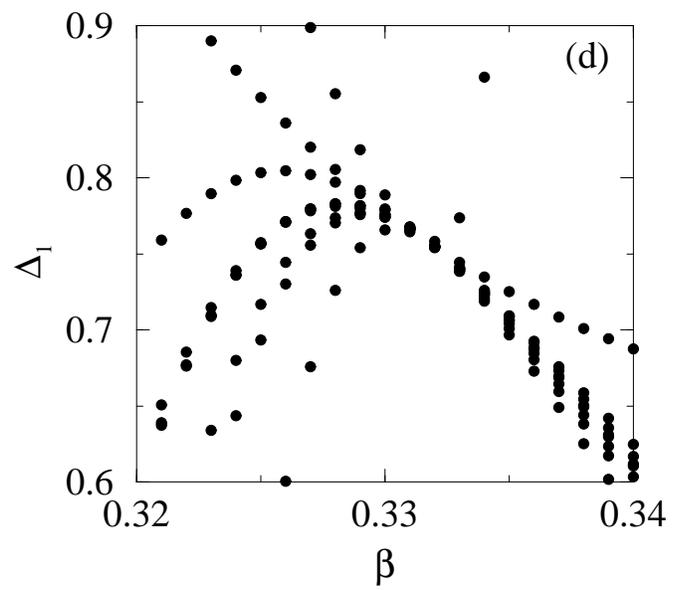
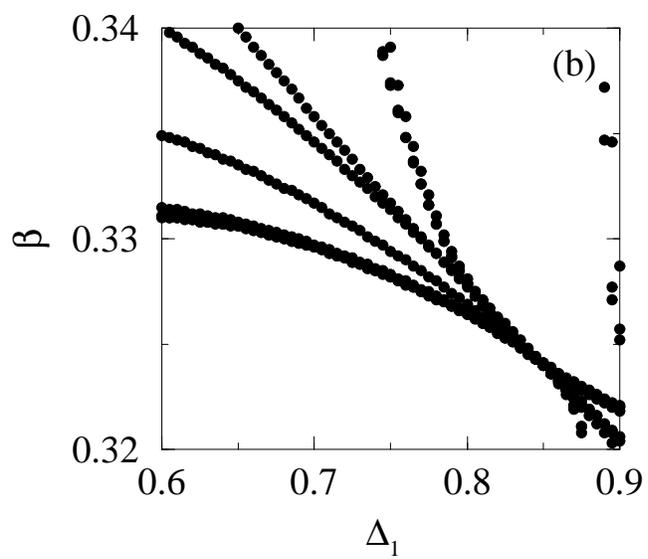
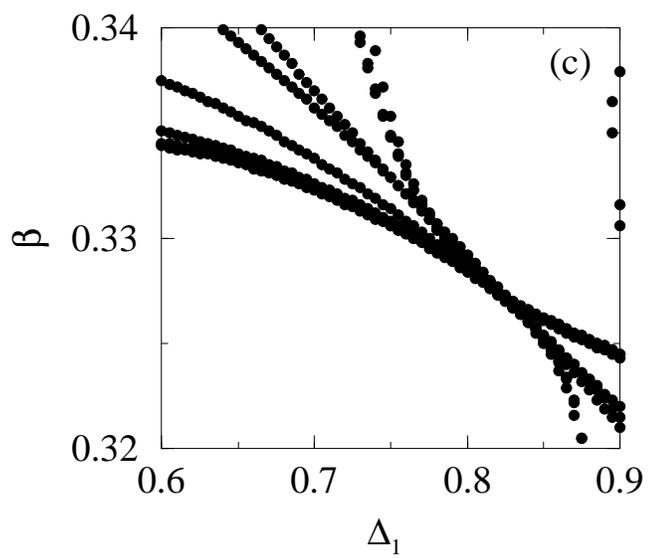
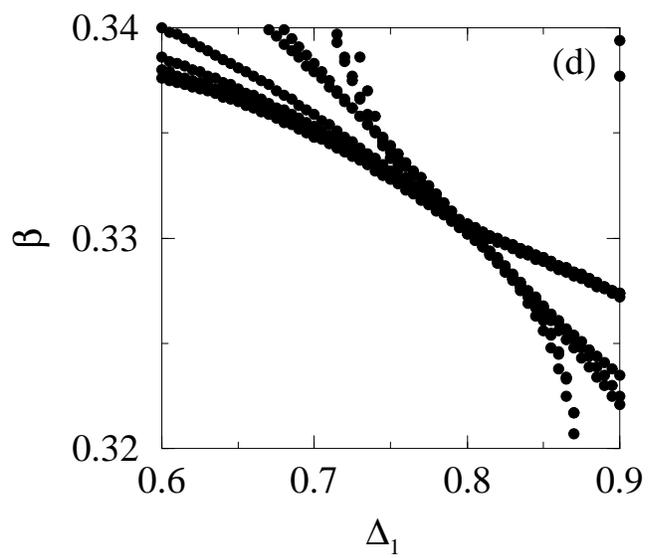
\end{document}